\documentclass[twocolumn,floatfix,prb,aps,showpacs]{revtex4-1}
\usepackage{graphicx,amsmath,amssymb,color}
\usepackage{color,amsmath,amssymb,amsfonts,graphicx,tabularx}
\usepackage{nicefrac}
\usepackage[titletoc,title]{appendix}
\usepackage[unicode=true,colorlinks=true]{hyperref}
\hypersetup{linkcolor=blue,citecolor=blue,urlcolor=blue}
\makeatletter
\newcommand*{\rom}[1]{\expandafter\@slowromancap\romannumeral #1@}
\makeatother

\newcommand{\be}{\begin{equation}}
\newcommand{\ee}{\end{equation}}

\newcommand{\ba}{\begin{eqnarray}}
\newcommand{\ea}{\end{eqnarray}}

\renewcommand{\vec}[1]{{\textbf{\textit{#1}}}}

\begin{document}

\title{Surprising robustness of particle-hole symmetry for composite fermion liquids}
\author{G. J. Sreejith$^1$, Yuhe Zhang$^2$ and J. K. Jain$^2$}
\affiliation{$^1$Indian Institute of Science Education and Research, Pune 411008, India}\affiliation{$^2$Department of Physics, 104 Davey Lab, Pennsylvania State University, University Park, Pennsylvania 16802, USA} 
   
\begin{abstract} 
We report on fixed phase diffusion Monte Carlo calculations that show that, even for a large amount of Landau level mixing, the energies of the Pfaffian and anti-Pfaffian phases remain very nearly the same, as also do the excitation gaps at $1/3$ and $2/3$. These results, combined with previous theoretical and experimental investigations, indicate that particle hole (PH) symmetry for composite fermion states is much more robust than a priori expected, emerging even in models that explicitly break PH symmetry.  We provide insight into this fact by showing that the low energy physics of a generic repulsive 3-body interaction is captured, to a large extent and over a range of filling factors, by a mean field  approximation that maps it into a PH symmetric 2-body interaction. This explains why Landau level mixing, which effectively generates such a generic 3-body interaction, is inefficient in breaking PH symmetry. As a byproduct, our results  provide a systematic construction of a 2-body interaction which produces, to a good approximation, the Pfaffian wave function as its ground state.
\end{abstract}
\maketitle

\section{Background and Motivation}

It has been appreciated since the early 1980s that Landau level (LL) mixing leads to corrections in various observable quantities in fractional quantum Hall effect (FQHE), such as the transport gaps\cite{MacDonald84,Melik-Alaverdian95}. The effect of LL mixing is two-fold: it alters the 2-body interaction and also induces multi-particle interaction. While both lead to corrections, the latter also breaks the symmetry under particle-hole (PH) transformation, which is an exact symmetry for electrons confined to a single LL and interacting by a 2-body interaction, providing an exact relation between the wave functions and spectra at filling factors $\nu$ and $1-\nu$. 

The issue of PH symmetry has received much attention in recent years, in the contexts of both the competition of the Moore-Read Pfaffian\cite{Moore91} (Pf) and the anti-Pfaffian\cite{Levin07,Lee07} (APf) states for the $\nu=5/2$ FQHE\cite{Bishara09,Wojs10,Rezayi11,Peterson13,Pakrouski15,Zaletel15,Rezayi17}, and the composite-fermion (CF) Fermi sea at $\nu=1/2$\cite{Son15,Geraedts15,Wang15,Balram16b,Wang16,Wang17}. A surprising message revealed by these studies is that the PH symmetry for CF liquids is much more robust than one might a priori anticipate, emerging even for Hamiltonians that explicitly break PH symmetry. Let us list some examples. (i) The most dominant PH symmetry breaking term induced by LL mixing is the 3-body interaction\cite{Bishara09}.  Ref.~\cite{Wojs10b} found, surprisingly, that even a \emph{pure} 3-body interaction produces, for a broad range of parameters, the standard Jain CF states at $\nu=n/(2n\pm 1)$\cite{Jain89,Jain07}, which satisfy PH symmetry to a very good approximation; this demonstrates an emergent PH symmetry in the ground state for a model Hamiltonian that explicitly breaks PH symmetry.  (ii) An emergent PH symmetry was found also for spinful bosons in the lowest LL (LLL) interacting via the hard core interaction\cite{Geraedts17}; a priori, a bosonic system is not expected to obey PH symmetry at all. (iii) Wang {\em et al.}\cite{Wang17} have demonstrated that the Halperin-Lee-Read  theory of composite fermions produces results that are consistent with PH symmetry, even though the theory, strictly speaking, breaks the PH symmetry because it is not confined to the LLL. (iv) It has proven surprisingly difficult to ascertain, theoretically, whether the Pf or the APf is selected by LL mixing\cite{Wojs10,Rezayi11,Peterson13,Pakrouski15,Zaletel15,Rezayi17}. Some theoretical calculations favor the Pf while others the APf, and a recent article\cite{Rezayi17} argues that the 3-body pseudopotential in the $m=9$ angular momentum channel is  the decisive factor, indicating the extreme subtlety of the issue.  (v) In addition, experiments also fail to find clear evidence for breaking of PH symmetry. In the experiments that measure the Fermi wave vector of composite fermions through commensurability oscillations, no deviation from the ``minority density rule"\cite{Kamburov14b} is seen even for hole type samples where LL mixing can be significant\cite{Jo17}.

Our goal in this work is two-fold. First, we search for signs of breaking of the PH symmetry due to LL mixing through the fixed phase diffusion Monte Carlo method\cite{Reynolds82,Foulkes01,Ortiz93,Melik-Alaverdian97,Zhang16}, which treats LL mixing in a non-perturbative fashion. We find that while the energy expectation value of both the Pf and the APf wave functions change substantially by LL mixing, their difference remains insignificant up to large LL mixing. Furthermore, we find that the transport gaps for the fully spin polarized states at $\nu=1/3$ and $\nu=2/3$ remain very nearly equal (in Coulomb units) even for large LL mixing. These results provide further support to the robustness of PH symmetry for CF liquids.

Our second goal is to seek insight into the surprising robustness of PH symmetry for the CF states. Because breaking of the PH symmetry can be modeled primarily as a 3-body interaction within the Hilbert space of a given LL, it is natural to consider a 3-body interaction to address the issue. We consider the validity of a mean field (MF) approximation that maps a 3-body interaction into a PH symmetric 2-body interaction. By explicit evaluation of the spectra for small systems, we find that the latter captures the physics of the 3-body interaction to a surprising degree, thus indicating that the 3-body interaction only weakly breaks PH symmetry.  Furthermore, the MF approximation improves as the 3-body interaction is made longer in range. This result gives insight into why LL mixing is very inefficient in breaking PH symmetry, because we expect LL mixing generically to produce a long-range 3-body interaction. As a byproduct, the MF mapping from 3-body to 2-body interaction also allows us to determine 2-body interaction for which the Pfaffian state is an accurate ground state.

\section{Fixed Phase Diffusion Monte Carlo Study}

Fixed-phase diffusion Monte Carlo (DMC) method \cite{Reynolds82,Foulkes01,Ortiz93,Melik-Alaverdian97,Zhang16} has proved successful in treating LL mixing non-perturbatively. This DMC method\cite{Reynolds82,Foulkes01} is based on the idea that the Schr\"odinger equation in imaginary time $\tau=it$ is equivalent to a diffusion equation, with the wave function representing the density of particles, and the long time limit of the evolution projects into the ground wave function. This produces, in principle, the exact ground state wave function provided it is everywhere real and non-negative. For electrons in a magnetic field, the wave function is necessarily complex, but Ortiz, Ceperley and Martin \cite{Ortiz93} developed a fixed phase DMC in which one expresses the wave function as $|\Phi(\vec{R})|e^{i\phi(\vec{R})}$ ($\vec{R}$ represents the particle coordinates collectively), assumes the phase to be fixed, which can be incorporated into the Hamiltonian, and then determines the lowest energy state within the chosen phase sector.  This method was generalized for spherical geometry by Melik-Alaverdian {\em et al.}~\cite{Melik-Alaverdian97, Melik-Alaverdian01}.

The accuracy of the fixed phase DMC method depends on the choice of phase $\phi(\vec{R})$. G\"u\c{c}l\"u and Umrigar~\cite{Guclu05} showed, by comparison with exact diagonalization, that LL mixing does not significantly alter the phase of LLL eigenstates. In our calculation, we therefore use the LLL projected wave functions for composite fermions as the initial trial wave functions to fix the phase in the DMC procedure. The validity of this approach has also been demonstrated by its ability to accurately capture the physics of spin polarization phase transitions in FQHE \cite{Zhang16}.

To evaluate the correction due to LL mixing at $\nu=5/2$ using DMC, we take as our initial trial wave functions the Pf and the APf states. We obtain the Slater determinant decomposition of a $1/2$ Pfaffian state through the Jack polynomial method ~\cite{Bernevig08, Bernevig09,Thomale11,Lee14}, which is an efficient way to determine the exact Slater expansion of certain FQHE model wave functions. The APf is obtained straightforwardly by PH conjugation. We construct the Pf and APf wave functions in the up-spin sector at $\nu=3/2$ by combining a fully filled LLL and forming the Pf/APf state in the second LL. Further multiplication by the $\nu=1$ state with down spin produces the Pf/APf trial states at $\nu=5/2$. We consider below both the fully spin polarized state at $\nu=3/2$ and the partially spin polarized state at $\nu=5/2$.
 
\begin{figure}
\includegraphics[width=0.95\columnwidth]{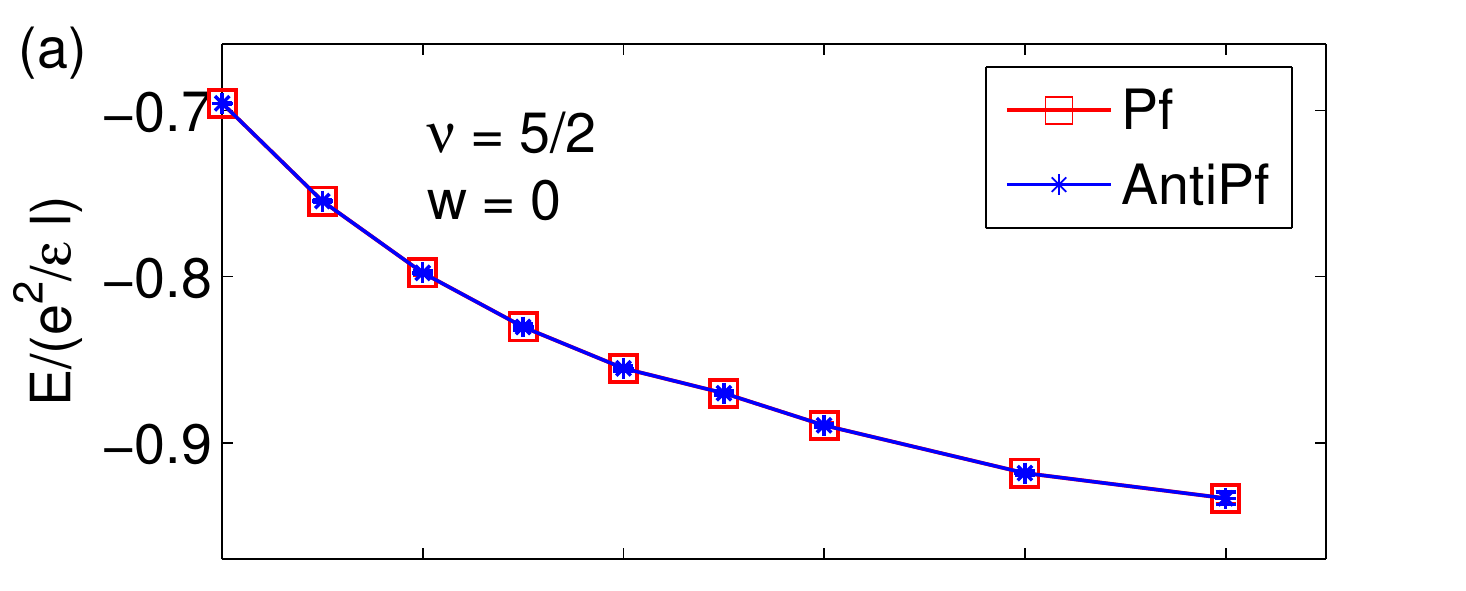}\\
\vspace{-2.1mm}
\includegraphics[width=0.95\columnwidth]{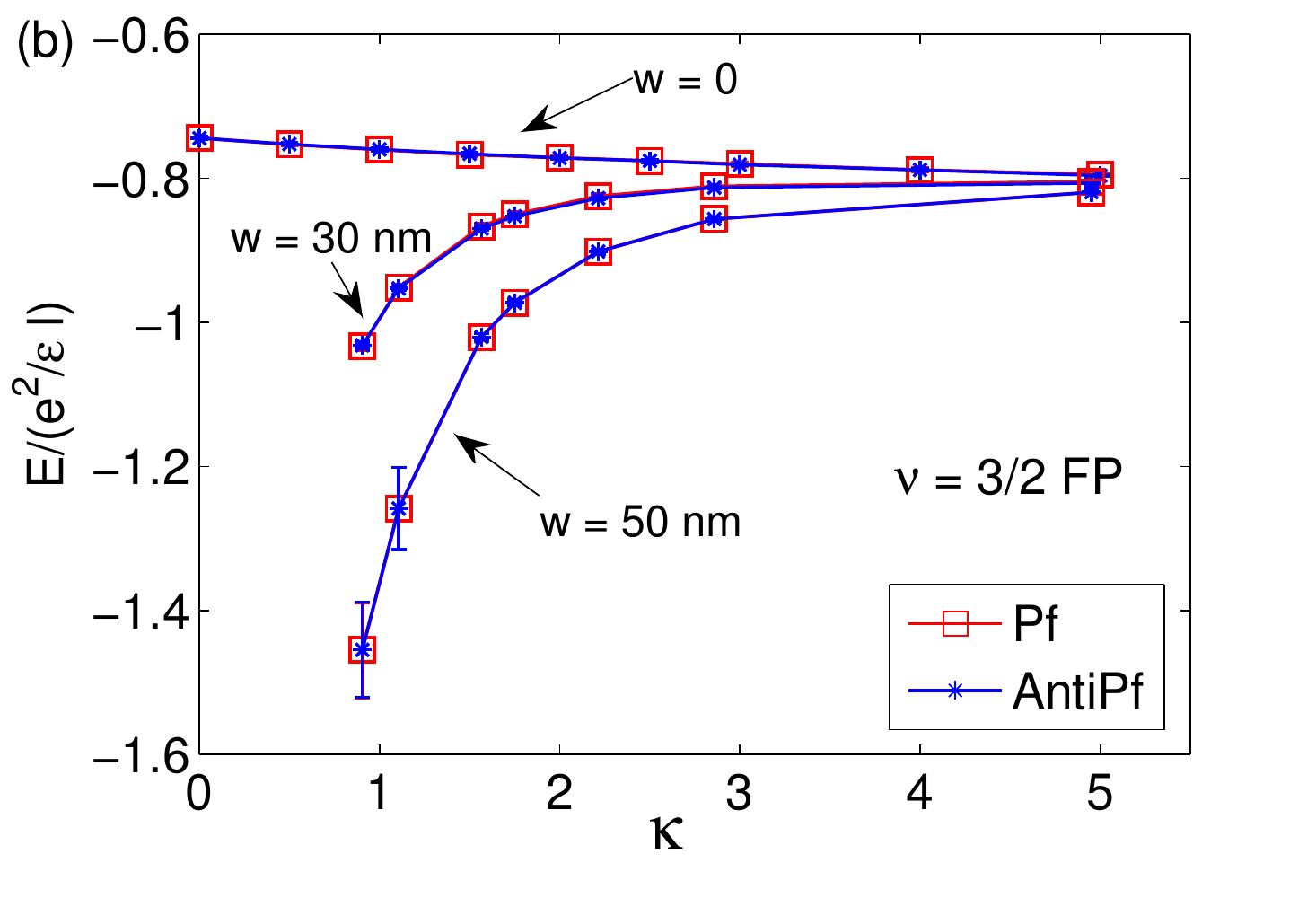}
\caption{(a) The DMC energies for Pfaffian and anti-Pfaffian states at $\nu = 5/2$ as a function of $\kappa$ for ideal 2D system ($w = 0$), obtained from extrapolation from systems with up to 32 and 30 particles, respectively. (b) The energies for spin polarized (SP) states at $\nu=3/2$ for $w=0$ and quantum wells of width $w = 30$nm and $50$nm (obtained from extrapolation of energies of up to 20 and 18 particles). All energies shown are obtained by extrapolating finite-system results to thermodynamic limit with the error bars indicating the extrapolation errors.
\label{pf}}
\end{figure}

To characterize the LL mixing, we use the parameter $\kappa = \hbar \omega_{c} / (e^{2}/ \epsilon l)$, namely the ratio of the cyclotron energy to the Coulomb energy. Fig.~\ref{pf} (a) shows the energies for Pfaffian and anti-Pfaffian states at $\nu=5/2$ in an ideal 2D system (with width $w=0$). The energies of the Pf and the APf states show a significant decrease due to LL mixing, dropping by more than 30\%  as $\kappa$ increases from 0 to 5. Fig.~\ref{pf} (b) shows the energies for fully polarized (FP) states at $\nu = 3/2$ for both $w = 0$ and for quantum wells with nonzero width.  For a quantum well (QW), we have used an effective 2D interaction based on the realistic charge distribution in the perpendicular direction (using the same method as in Ref. \onlinecite{Zhang16}; see references therein for further details).  The energies at finite width increase with increasing $\kappa$, which may seem counter-intuitive but is an effect of finite width which depends on the density; the effective interaction softens with increasing density, and $\kappa$ is reversely related to areal density $\rho$ as $\kappa \approx 1.28\sqrt{\nu/(\rho/10^{11}\text{cm}^{-2})}$. In all cases, the Pfaffian and anti-Pfaffian states behave indistinguishably with LL mixing within our numerical uncertainty. As an interesting aside, we note that the inclusion of the spin-down LL substantially increases the correction due to LL mixing, as can be seen by comparing the $w=0$ energies in the two panels of Fig.~\ref{pf}.

\begin{figure}
\includegraphics[width=0.9\columnwidth]{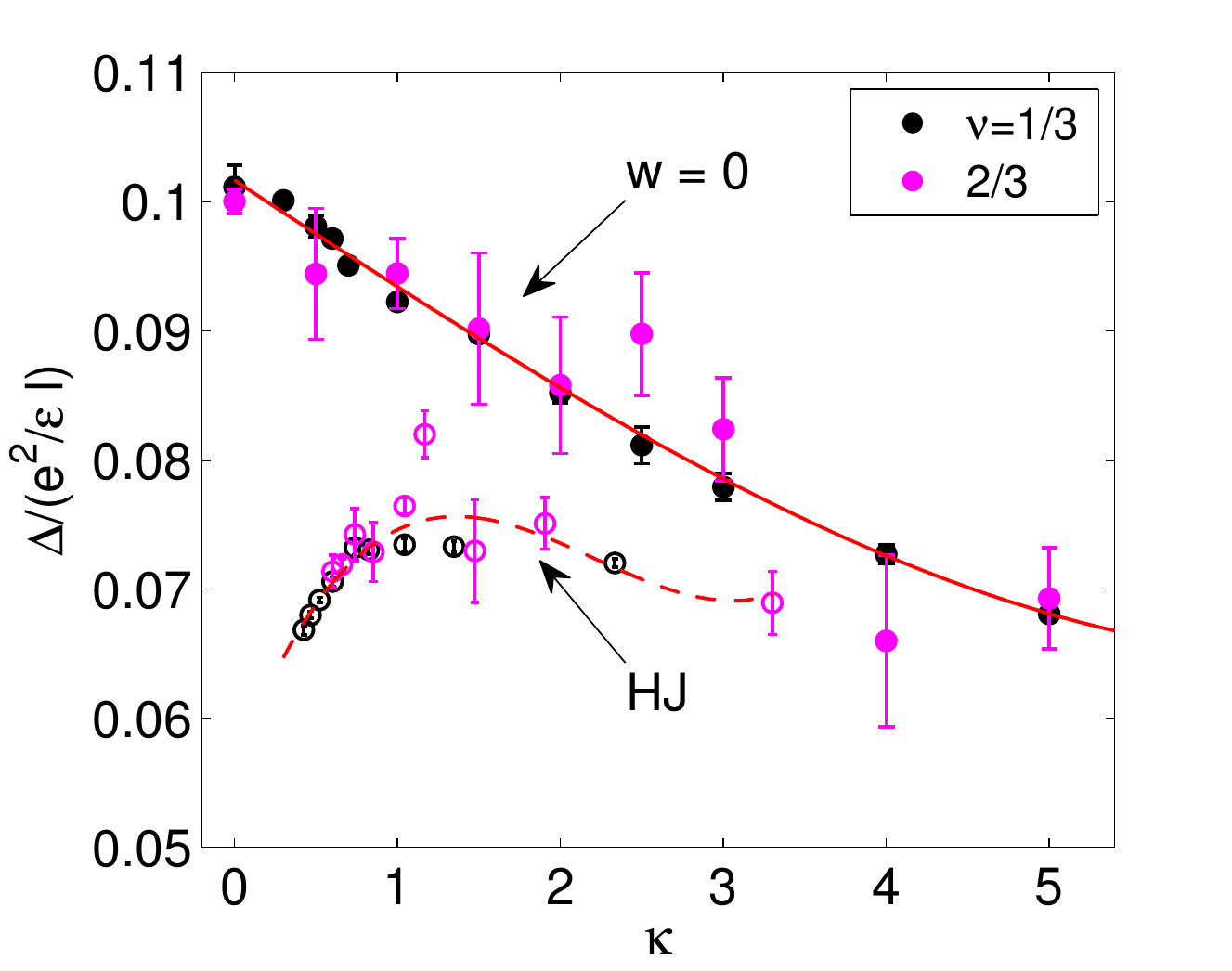}
\caption{The energy gaps of 1/3 and 2/3 states as a function of $\kappa$ for different filling factors for $w = 0$ (solid symbols) and heterojunction (HJ - empty symbols). These are thermodynamic extrapolations from gaps of systems with up to 32 and 20 particles. Solid and dashed lines are a guide to the eye.
\label{gap}}
\end{figure}

As another test of PH symmetry breaking, we consider the excitation gaps of FQH states at filling factors $\nu$ and $1 - \nu$; these should be equal in units of $e^2/\epsilon l$ when LL mixing is absent. The excitation gap measured in transport experiments (as the activation energy deduced from the Arrhenius behavior of the longitudinal resistance) corresponds to the energy required to create a far-separated CF-particle-CF-quasihole pair. We directly use the wave function of the CF-exciton at large wave vector limit as trial wave function to perform DMC calculation. Fig.~\ref{gap} shows that the gaps for $\nu = 1/3$ and 2/3 are reduced by up to 25\% due to LL mixing as $\kappa$ is changed from 0 to 5. However, the two gaps remain indistinguishable within numerical uncertainty, indicating that the PH symmetry is preserved, to a good approximation, with LL mixing.

\section{MF approximation for the 3-body interaction}

As mentioned above, when the Hamiltonian is projected into a given LL, the effect of LL mixing manifests through correction to the 2-body interaction and appearance of 3- and higher body interaction terms. The latter are responsible for breaking the PH symmetry. We now ask how efficient the 3-body interaction is in breaking the PH symmetry. In particular, we will consider a MF approximation for a 3-body interaction wherein it is mapped into a 2-body interaction. If this approximation is accurate, we can conclude that the generic 3-body interaction preserves PH symmetry to a good approximation. We shall assume a fully spin polarized state below.

In order to study the bulk physics of the many electron systems, we will use the spherical geometry, wherein the electrons move on the surface of a sphere exposed to $2Q$ magnetic flux quanta (where a flux quantum is $\phi_0=hc/e$). Here a general 3-body interaction can be written as 
\begin{equation}
\mathcal{V}^{(3)} = \frac{1}{3!3!}\sum_{\{q_i;k_i\}} V^{(3)}_{q_1,q_2,q_3;k_1,k_2,k_3} c^\dagger_{q_3} c^\dagger_{q_2} c^\dagger_{q_1} c_{k_1} c_{k_2} c_{k_3}
\end{equation}
where the indices $q_i,k_i$ correspond to the $L_z$ quantum numbers of the electrons each of which has a total orbital angular momentum $Q$, equal to half the number of flux quanta passing through the surface of the sphere. For a rotationally symmetric system, the matrix elements $V_{q_i;k_i}$ can be expanded in terms of the projectors $P^{(3)}_l$ into the angular momentum $l$ subspace of three electrons.
\begin{equation}
\mathcal{V}^{(3)} = \sum_{l=3}^{2Q} V^{(3)}_{l} P^{(3)}_{3Q-l}
\end{equation}
where the 3-body Haldane pseudopotential \cite{Haldane83,Simon07a} $V^{(3)}_l$ 
is the energy of three electrons in the state with total angular momentum $3Q-l$. (There can be several independent states with the same $l,l_z$ quantum numbers for $l>5$, but we will not 
consider those in the present article.) In what follows, we will consider the 3-body model with $V^{(3)}_3\equiv A$, $V^{(3)}_5\equiv B$, and $V^{(3)}_l\equiv 0$ for $l>5$. (Note that there are no states with $l=0, 1, 2$ and 4.)

\begin{figure}
\includegraphics[width=\columnwidth]{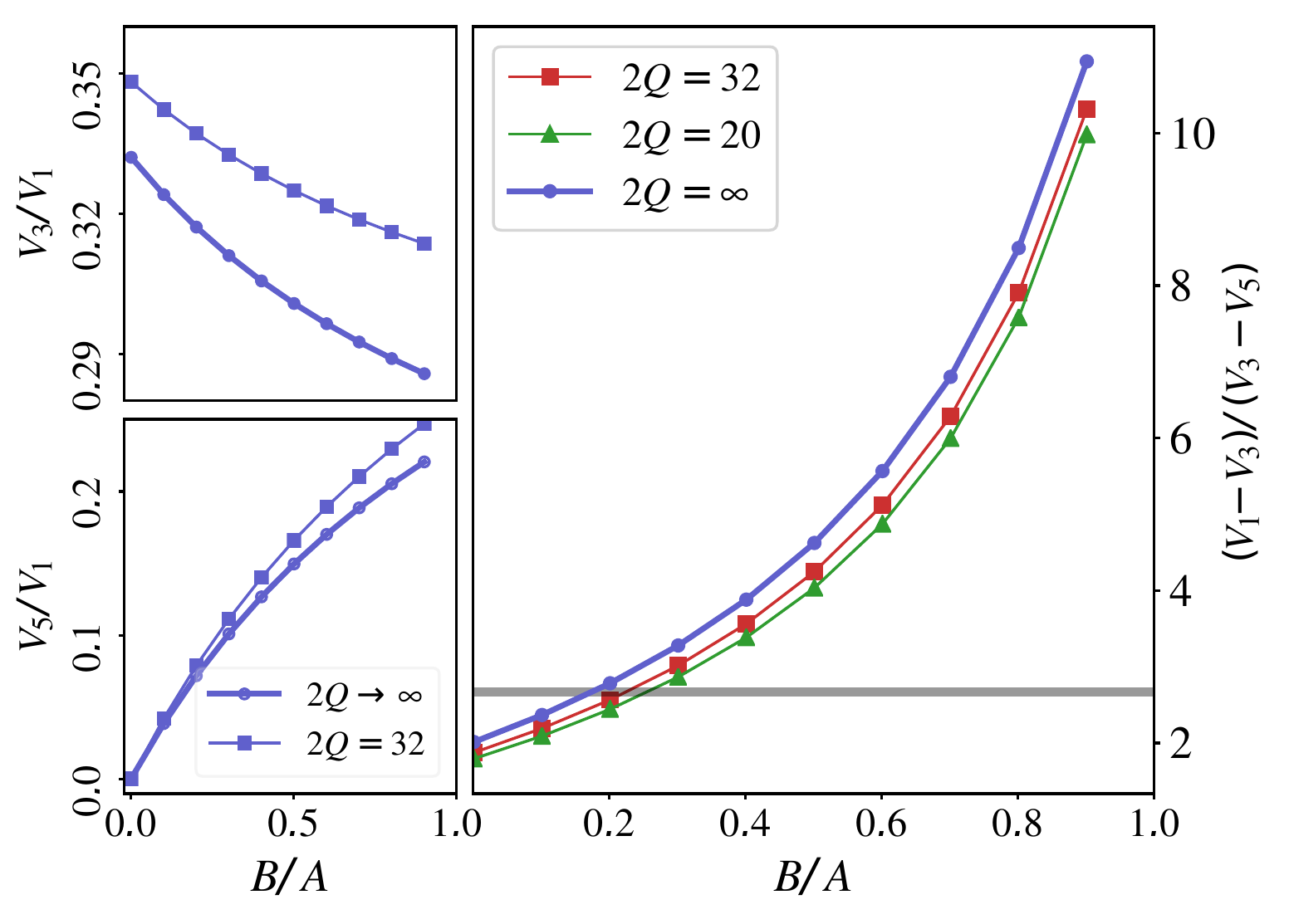}
\caption{(left): Pseudopotentials of the MF approximation to the 3-body interaction for the case where only the first two 3-body pseudopotentials ($A,B$) are non-zero. (right): Ratio $(V_1-V_3)/(V_3-V_5)$ as a function of $B/A$. The dotted horizontal line shows the ratio for the case of LLL Coulomb interaction $2.67$. Mean field Hamiltonian reproduces this ratio at $B=0.18$ in the the large $Q$ limit and $B>0.18$ in finite $Q$ systems.
The values of $(V_1-V_3)/(V_3-V_5)$ in the thermodynamic limit $Q=\infty$ were obtained by linearly extrapolating $V_l$ as a function of $1/Q$, and they coincide with the pseudopotentials in the flat geometry.
\label{FigA1B}}
\end{figure}

We propose to approximate
 this 3-body interaction by a simpler 2-body Hamiltonian in which we replace one factor $c^\dagger_{q_1}c_{k_1}$ with its ground state expectation values. Since we are primarily interested in exploring homogeneous, rotationally symmetric ground states, these expectation values take the form $\left\langle c^\dagger_{q_1}c_{k_1} \right \rangle = \nu\,\delta_{q_1,k_1}$. This MF ansatz leads to a 2-body Hamiltonian of the form 
\begin{equation}
\mathcal{V}^{(2)}=\frac{1}{2!2!}\sum_{q_{1},q_{2};k_{1},k_{2}}V_{q_{1},q_{2};k_{1};k_{2}}c_{q_{1}}^{\dagger}c_{q_{2}}^{\dagger}c_{k_{1}}c_{k_{2}}
\end{equation}
where the matrix elements are given by the partial trace of the 3-body interaction Hamiltonian
\begin{equation}
V_{q_{1},q_{2};k_{1};k_{2}}=\nu\sum_{l=-Q}^Q\left\langle q_{1},q_{2},l\right|\mathcal{V}^{(3)}\left|k_{1},k_{2},l\right\rangle.
\end{equation}
It can be checked numerically that this results in a rotationally symmetric Hamiltonian whose 2-body pseudopotentials $V_l$ can be obtained by diagonalizing $\mathcal{V}^{(2)}$ for a two particle system to get:
\begin{equation}
\mathcal{V}^{(2)} = \sum_{l=2}^{2Q-1} V_l P^{(2)}_{2Q-l}
\end{equation}
Figure \ref{FigA1B} (left panels) show the 2-body MF pseudopotentials for $A=1$ and $B\geq 0$. Only the pseudopotentials in the $l=1,3$ and $5$ channels are non-zero and $V_{l=5}$ increases from $0$ with increase in $B$. (The 2-body pseudopotentials for even $l$ are not relevant for fully spin polarized states.) 

Note that the pseudopotentials of the MF Hamiltonian depend on $Q$. As the radius of the sphere scales with $Q$, in the limit of $Q\to\infty$, the interactions described on a sphere approach that of a flat geometry.  As shown in the Appendix \ref{App-Disc}, in the disk geometry, we find 
$V^{(2)}_1= 3.375A\nu+2.53125B\nu$, $V^{(2)}_3 = 1.125A\nu+0.5625B\nu$, and 
$V^{(2)}_5 = 1.40625 B\nu$, which give precisely the $2Q\rightarrow\infty$ limits of the MF pseudopotentials on the sphere. In what follows, we will consider the finite $2Q$ pseudopotentials of the spherical geometry.

One may ask when the 3-body interaction most resembles the 2-body Coulomb interaction. For this purpose, it is convenient to define the ratio $(V_1-V_3)/(V_3-V_5)$; this ratio is an appropriate characterization of the short range behavior of a given 2-body interaction because the eigenstates are invariant under a change of scale as well as an addition of a constant to all $V_l$. For the ideal Coulomb interaction in the LLL, this ratio is 2.67.  This ratio for the MF interaction is shown in the right panel of Fig.~\ref{FigA1B}. We see that the MF approximation on the 3-body interaction given by $(A,B)=(1,0.18)$ best represents the Coulomb interaction. Because one expects the standard CF physics for 2-body interactions that are Coulomb like or more repulsive, we expect that, provided that the MF approximation is valid, the CF physics should occur for a pure 3-body interaction for $B/A \gtrapprox 0.18$.

\section{Testing the MF theory}

We now compare the eigenspectra and eigenstates of the MF Hamiltonian and the 3-body interaction at several specific flux values. For our 3-body interaction model, the Pfaffian state at $\nu=1/2$ is the exact zero energy ground state for $B=0$ \cite{Greiter92a} and the gaffnian state at $\nu=2/5$ is the exact ground state when both $A$ and $B$ are non-zero\cite{Simon07b}. The spectrum of these interactions for $\nu<1/2$ and $\nu<2/5$, respectively, contains a very large number of zero-energy eigenstates. Therefore, we will consider only $\nu>2/5$, because for filling factor $\nu\leq 2/5$, it would be necessary to switch on further 3-body pseudopotentials to model a generic 3-body interaction.

\begin{figure}
\includegraphics[width=\columnwidth]{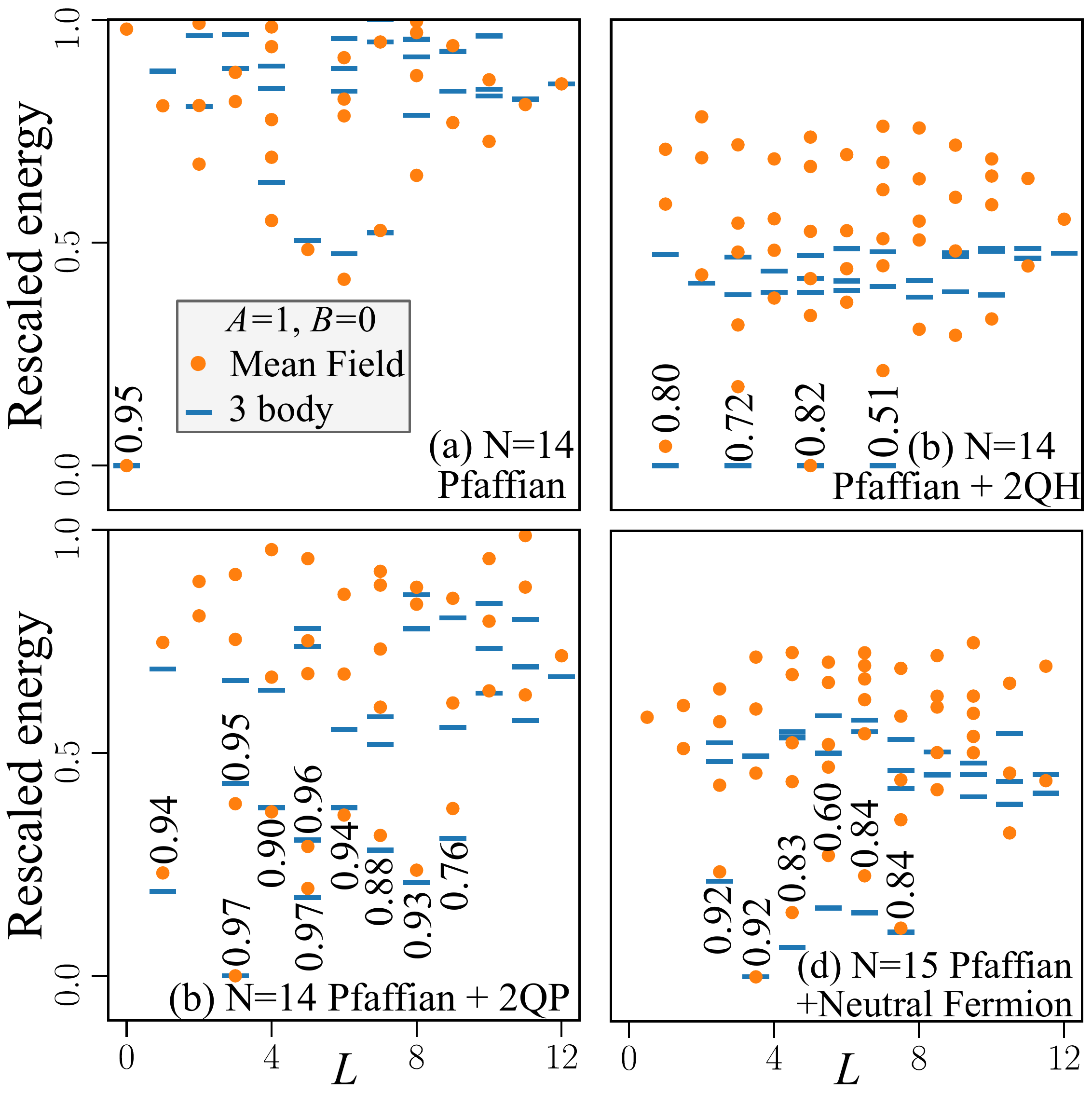}
\caption{Comparison of the spectra of the 3-body interaction $(A,B)=(1,0)$ for the $\nu=1/2$ Pfaffian state and its excitations (blue dashes) with those of the corresponding 2-body mean-field interaction (orange dots). The numbers next to the low energy states show the overlap between the lowest energy states of the two interactions at the each angular momenta. Panel (a), (b) and (c) show the spectra for 
$2Q=25$, $26$ and $24$ for $N=14$ particles, while (d) for $2Q=27$ for $N=15$. 
The spectra have been shifted vertically to align the ground states, and the MF energies have been divided by a factor of $1.5$.
\label{N14PfaffianB0}}
\end{figure}

We first consider the 3-body interaction $(A,B)=(1,0)$. The 3-body spectrum in Fig.~\ref{N14PfaffianB0}a contains the Pfaffian as the exact zero energy state and a collective mode of neutral excitations\cite{Sreejith11a,Sreejith11b}. The corresponding 2-body MF interaction reproduces a remarkably similar spectrum, with (i) a ground state with a high overlap with the Pfaffian; (ii) a neutral mode with the same counting and dispersion; and (iii) gaps accurate up to a factor of $1.5$. Removal of a single flux from this system produces two quasiparticles of the Pfaffian (Fig.~\ref{N14PfaffianB0}(c)); the MF approximation again nicely reproduces the structure, and its eigenstates have high overlaps with those of the $(A.B)=(1,0)$ model. For $2Q=2N-3$ with an odd $N$, the 3-body interaction spectrum contains a topological exciton\cite{Sreejith11a,Sreejith11b} (also called a neutral fermion), which is also closely reproduced by the MF interaction (Fig.~\ref{N14PfaffianB0}d). Finally, the state $2Q=2N-2$ has two quasiholes that form a band of exact zero energy states, as seen in Fig.~\ref{N14PfaffianB0}(c). As noted above, the 3-body interaction with $B=0$ is non-generic here in that it produces zero energy states, but still the MF Hamiltonian produces a low energy band with states at the same quantum numbers.

\begin{figure}
\includegraphics[width=.95\columnwidth]{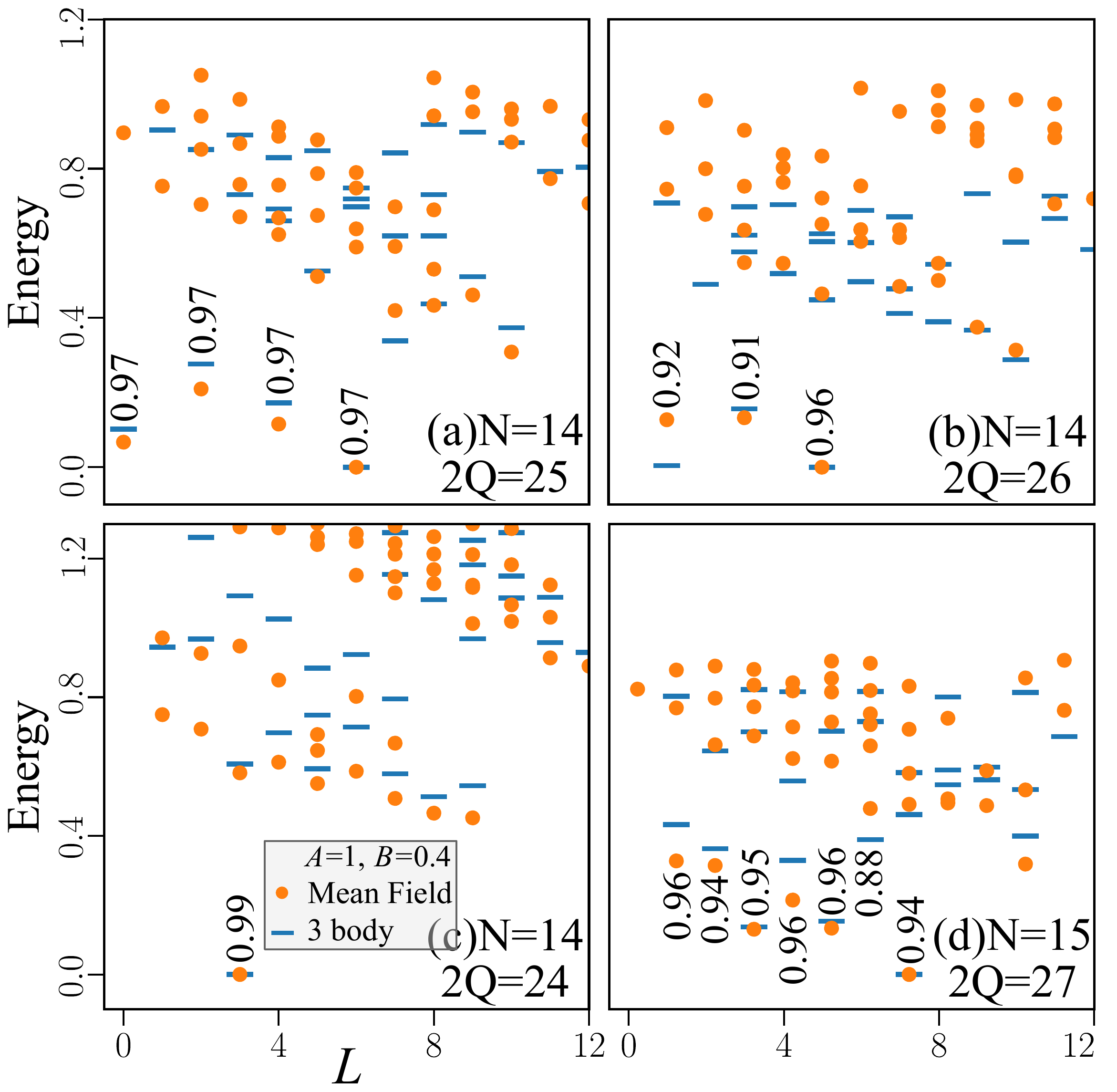}
\caption{Comparison similar to that in Fig.~\ref{N14PfaffianB0} at the same particle numbers and flux values, but for interaction parameters $B=0.4$. The MF interaction continues to be a good approximation to the 3-body interaction. The spectrum here resembles that of the LLL Coulomb interaction at the corresponding particle numbers and fluxes. The low energy states here can be identified with particle hole conjugate of composite fermion states with (a) two quasiparticles of $2/5$, (b) two quasiholes of $3/7$, (c) single quasihole of $2/5$ and (d) three quasiparticles of $2/5$. The spectra have been shifted vertically to align the ground states, but without any rescaling of the energies.
\label{N14PfaffianB4}}
\end{figure}

\begin{figure}[h!]
\includegraphics[width=\columnwidth]{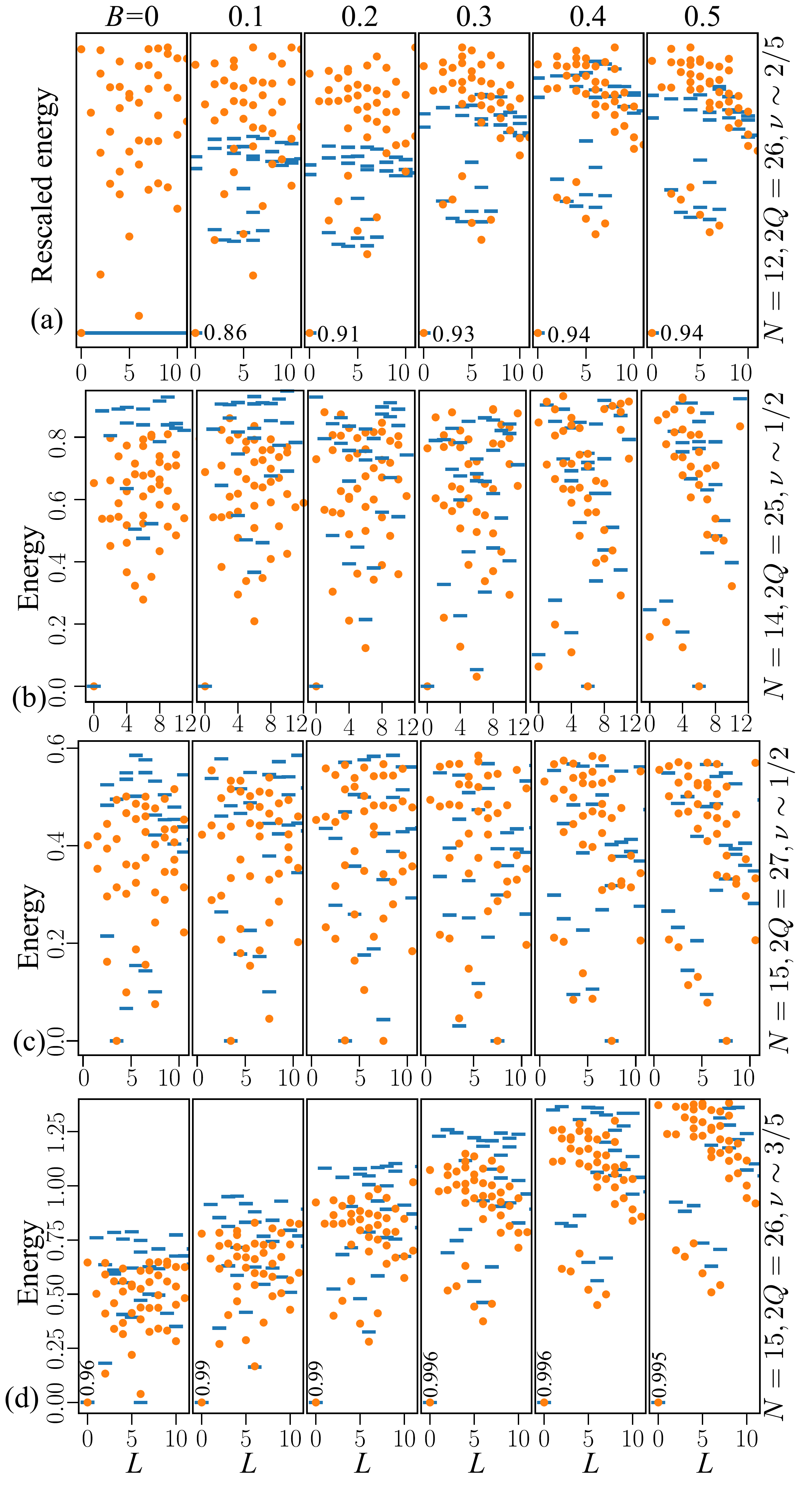}
\caption{Evolution of the low energy spectrum in the flux sector corresponding to the CF state at $\nu=2/5$ (a), Pfaffian state (b), Pfaffian state with a topological exciton (c) and the CF state at $\nu=3/5$ (d) as $l=5$ pseudopotential $V^{(3)}_5\equiv B$ is increased. Blue dashes show the spectra for the 3-body interaction while the orange dots show the spectra for the 2-body mean field interaction. Energies have been shifted vertically to set the ground states at zero energy. Energy has been rescaled by a $B$ dependent factor for alignment in the case of $2/5$ but not in the others. The three body spectrum at $\nu=2/5, B=0$ (a) contains a large number of zero energy states as a result of which the spectrum appears like a single line at $E=0$.
\label{Bseq}}
\end{figure}

Fig.~\ref{N14PfaffianB4} shows a similar comparison for $(A,B)=(1,0.4)$.
The low energy spectra are qualitatively different from those in Fig.~\ref{N14PfaffianB0}, and are similar to the spectrum of the LLL Coulomb interaction. The counting of the low energy spectrum matches exactly with the prediction of the composite fermion theory (see Ref \cite{Wojs10b}). 

We stress that the MF theory produces not only accurate eigenstates but also the energy scales quite well without any rescaling. We find this to be the case for $B>0$ and for filling fraction away from (and greater than) $\nu=2/5$, as seen in the comparisons below.

The MF Hamiltonian works well at flux values corresponding to other filling fractions in the range $\nu>2/5$. For example, Fig.~\ref{Bseq} (d) shows that it successfully reproduces, even for $B=0$, the 3-body spectra at a flux where the ground state $\nu=3/5$ state occurs in the LLL.  The panel (b) of Fig-\ref{Bseq} shows the evolution of the 3-body and the MF spectra as a function of $B$ for the flux corresponding to the Pfaffian state. The energy scales are captured well by the MF spectra except for small $B$. Similarly panel (c) shows the evolution of the spectra for the flux at which we find a topological exciton of the Pfaffian. At large $B$, the spectrum corresponds to that of three quasiholes of the $3/5$ state.

As noted above, we do not expect the MF approximation to be valid for $\nu< 2/5$, where the $(A,B)$ model produces zero-energy non-Abelian quasiholes of the Gaffnian 2/5 state, the physics of which can obviously not be captured by a 2-body interaction. The case of $\nu=2/5$ is interesting (Fig-\ref{Bseq}(top)). Here the 3-body interaction produces a spectrum, at finite $B$, that is similar to the spectrum at the PH conjugate filling $\nu=3/5$ (panels (a) and (d) of Fig-\ref{Bseq}), but with a much smaller gap. The MF approximation, being PH symmetric, produces exactly the same spectrum as at $3/5$, but with the energy scale reduced by a factor of 2/3 due to the factor of $\nu$ in the pseudopotential. The energies of the neutral excitations in the 2-body MF spectra are nonetheless much larger than those in the 3-body spectra. 
We believe this is a consequence of the fact that the 2-body interaction does not contain the physics of zero energy quasiholes of the 3-body interaction.

\section{Discussion}

Success of the MF theory gives insight into why the 3-body interaction produces the standard CF physics at nonzero $B$. For a range of parameters, the 3-body interaction corresponds to a MF interaction that has a sufficiently strong short range repulsion. 
As known from previous studies, 2-body interactions with sufficiently strong short-range repulsion produce composite fermions carrying two vortices. This explains why the 3-body interaction produces, except for very small $B$, the standard CF-Fermi sea like spectrum at $\nu=1/2$ (Fig.~\ref{N14PfaffianB4}), and the standard lowest-LL state at $\nu=3/5$ (Fig.~\ref{Bseq}) and other filling factors~\cite{Wojs10b}.

These considerations clarify why the 3-body interaction produces eigenstates that are very close to being PH symmetric. The main point is that the not all of a given 3-body interaction breaks PH symmetry. It is only the difference between the 3-body and the corresponding 2-body MF interaction that breaks PH symmetry. The fact that the 2-body MF interaction produces eigenstates and eigenenergies that are very close to eigenstates and eigenenergies of the 3-body interaction indicates that the difference is small. 

Furthermore, we find that the MF interaction represents a better approximation when the 3-body interaction is long ranged, for example, when $B$ is nonzero. LL mixing is expected generically to produce a 3-body interaction for which $V^{(3)}_l$ are all nonzero and decay monotonically with $l$. We expect that such an interaction is well represented by the 2-body MF interaction, and thus only weakly breaks PH symmetry. This explains why our fixed phase DMC calculations are not able discriminate between the Pf and the APf states, or why experiments do not see a clear signature of breaking of PH symmetry even when the LL mixing parameter $\kappa$ is very large.

\section{Acknowledgments}  

The work at Penn State was supported in part by the US Department of Energy under Grant No. DE-SC0005042. The authors gratefully acknowledge the Institute for CyberScience at the Pennsylvania State University and MPI-PKS, Dresden for providing computing resources. JKJ thanks Mohit Randeria for an insightful remark.

\appendix

\section{MF approximation in the disk geometry}
\label{App-Disc}
The generalized rotationally symmetric 3-body interaction in the disk geometry is such that there is a finite energy cost $V_l$ for any three particles to be in a relative angular momentum state $l$ (for fermions $l=3,5\dots$). Thus the Hamiltonian has the form 
\begin{equation}
{\mathcal V}^{(3)} = \sum_{l=0}^\infty V^{(3)}_l P_l^{(3)}
\end{equation}
The $P_l^{(3)}$ is a projector on to the relative angular momentum $l$ subspace of three particles and $V^{(3)}_l$ are the pseudopotentials. 

The relative angular momentum subspaces are highly degenerate, with different degenerate states further labelled by the center of mass angular momentum of three particles $T=0,1,\dots \infty$. For $l>5$ there are multiple states with same $(l,T)$ which calls for an additional label. This will not concern us in this work as we shall consider only those interactions for which $V^{(3)}_{l}=0$ for $l>5$. Thus for our purposes
\begin{multline}
{\mathcal V}^{(3)} = A \sum_{T=0}^{\infty}\left \vert l=3,T \right  \rangle  
\left \langle l=3,T \right  \vert + \\
B \sum_{T=0}^{\infty}\left \vert l=5,T \right  \rangle  
\left \langle l=5,T \right  \vert\label{discH}
\end{multline}
where the pseudopotentials $A=V^{(3)}_3$ and $B=V^{(3)}_5$ parameterize the interactions.

In terms of the single particle angular momentum states, the interaction can be written as follows:
\begin{equation}
{\mathcal V}^{(3)} = \sum_{{\bf p},{\bf q}}\left | \bf p\right \rangle
V^{(3)}_{{\bf p},{\bf q}}
\left \langle \bf q \right | 
\end{equation}
where
\begin{equation}
V^{(3)}_{{\bf p},{\bf q}} = \sum_{T=0}^\infty A\bar{\psi}^{l=3,T}_{\bf p}
\psi^{l=3,T}_{\bf q}
+
B  \bar{\psi}^{l=5,T}_{\bf p}
\psi^{l=5,T}_{\bf q}
\end{equation}
where $\left | {\bf p}\right\rangle = \left | p_1,p_2,p_3\right\rangle$ is the Slater determinant state of three fermions angular momenta $p_1,p_2$ and $p_3$,  ($p_1<p_2<p_3$). 
$\psi^{l,T}_{\bf p}$ represents the coefficients in the expansion 
\begin{equation}
\left|l,T\right \rangle = \sum_{\bf p} \psi^{l,T}_{\bf p}\left|\bf p\right \rangle
\end{equation}
Normalized states $\psi^{l,T}_{\bf p}$ can be obtained using exact diagonalization of a 2-body interaction with rotation and translation symmetry such as the Coulomb interaction in the space of three particles with total angular momentum $l+T$.

The interaction in the many particle system is given by
\begin{equation}
{\mathcal V}^{(3)} = \sum_{p_i,q_i=0}^\infty c^\dagger_{p_3} c^\dagger_{p_2} c^\dagger_{p_1} \frac{V^{(3)}_{{\bf p},{\bf q}}}{3! 3!}
c_{q_1} c_{q_2} c_{q_3}
\end{equation}
where $V^{(3)}_{{\bf p},{\bf q}}$ is taken to be antisymmetric within the $p_i$  and $q_i$ indices. The MF approximation involves substituting  $c_{i}^\dagger c_{j} \to \nu\delta_{ij}$ for one pair of operators in the above expression. This gives
\begin{equation}
{\mathcal V}^{(2)} = \sum_{p_i,q_i} c^\dagger_{p_2}c^\dagger_{p_1} \frac{V^{(2)}_{p_1,p_2;q_1,q_2}}{2!2!} c_{q_1}c_{q_2}
\label{DiscH2}
\end{equation}
where $V^{(2)}$ is given by the partial trace over one pair of indices
\begin{equation}
V^{(2)}_{p_1,p_2;q_1,q_2} = \nu
\sum_{p_3,q_3=0}^{\infty} \delta_{p_3q_3} V^{(3)}_{{\bf p},{\bf q}}
\end{equation}
The pseudopotentials $V^{(2)}_l$ of the MF Hamiltonian Eq (\ref{DiscH2}) can now be calculated by diagonalizing a system of two fermions in this interaction.

The MF approximation that maps ${\mathcal V}^{(3)}$ to ${\mathcal V}^{(2)}$ is linear. As a result the MF pseudopotentials $V^{(2)}_l(A,B)$ corresponding to the interaction $V^{(3)}(A,B)$ are simply the linear combination $AV^{(2)}_l(1,0)+BV^{(2)}_l(0,1)$
where $V^{(2)}_l(1,0)$ and $V^{(2)}_l(0,1)$ are given in the accompanying table.
\begin{table}[h!]
\begin{tabular}{|c|c|c|}
\hline 
 & $A=1,B=0$ & $A=0,B=1$\\
\hline 
$\overline{V}_{1}$ & $3.375\nu$ & $2.53125\nu$\\
\hline 
$\overline{V}_{3}$ & $1.125\nu$ & $0.5625\nu$\\
\hline 
$\overline{V}_{5}$ & $0$ & $1.40625\nu$\\
\hline 
\end{tabular}
\end{table}

Disc geometry can be pictured as $Q\to\infty$ limit of the spherical geometry wherein the relative and center of mass angular momentum quantum numbers correspond to the $L^2$ and $L_z$ quantum numbers in the spherical geometry. We indeed find that the MF pseudopotentials obtained in the large $Q$ limit (by extrapolation of the pseudopotentials calculated at finite $Q$) on the sphere match the ones for the disk geometry.

\bibliography{meanfield.bib}

\end{document}